\documentclass[prl,twocolumn,groupedaddress,showpacs,citeautoscript,floatfix]{revtex4}
\usepackage{graphicx}

\newcommand{\lbar}{\ensuremath{\,{\mathchar'26\mkern-8mul}}}

\begin{document}


\title{A continuous isotropic-nematic liquid crystalline
transition of F-actin solutions}


\author{Jorge Viamontes}
\author{Jay X. Tang}
\email[E-mail:]{jxtang@indiana.edu}
\affiliation{Physics Department, Indiana University, 727 East
Third St, Bloomington, IN 47405, U.S.A.}


\date{\today}

\begin{abstract}
The phase transition from the isotropic (I) to nematic (N) liquid
crystalline suspension of F-actin of average length $3~\mu$m or
above was studied by local measurements of optical birefringence
and protein concentration. Both parameters were detected to be
continuous in the transition region, suggesting that the I-N
transition is higher than 1st order. This finding is consistent
with a recent theory by Lammert, Rokhsar \& Toner (PRL, 1993,
70:1650), predicting that the I-N transition may become
continuous due to suppression of disclinations. Indeed, few line
defects occur in the aligned phase of F-actin. Individual
filaments in solutions of a few mg/ml F-actin undergo fast
translational diffusion along the filament axis, whereas both
lateral and rotational diffusions are suppressed.
\end{abstract}
\pacs{61.30.Eb, 61.30.Jf, 64.70.Md, 82.35.Pq, 87.15.-v}

\maketitle

The isotropic (I) to nematic (N) liquid crystalline transition
occurs in solutions of rodlike particles including stiff polymers
\cite{Sato97}, protein filaments \cite{Buxbaum87,Kerst90}, and
filamentous viruses \cite{Tang95}. Such a phase transition is
predicted to be 1st order based on statistical mechanical theories
\cite{Onsager49,Flory69}. Landau and de Gennes \cite{deGennes94}
treated the I-N transition by a general method of free energy
expansion, as a function of the nematic order parameter, which
contains a 3rd order term and thus predicts the I-N transition to
be 1st order. This classic analysis led to the conventional wisdom
that the I-N transition belongs to a different universality class
as the magnetic systems, for which the order-disorder transition
is generally continuous.

In a recent theory examining topology and nematic ordering by
Lammert, Rokhsar \& Toner (LRT) \cite{Lammert93}, it is shown,
however, that the weakly 1st order I-N transition can break into
two continuous transitions if the disclination core energy is
raised sufficiently high. The LRT theory has been developed with
the thermotropic nematogens in mind, although no particular
experimental system has been shown to date in direct support of
the theory. In this report, we first present experimental features
of the I-N transition of F-actin, and then discuss the relevance
of our findings to the theory of LRT.

The protein filaments F-actin provide us a challenging system to
study the liquid crystalline transition. On one hand, since the
protein is well characterized biochemically, special techniques
such as fluorescence labeling are available to probe both
structure and dynamics of an F- actin solution \cite{Kas95}. On
the other hand, extreme polydispersity in filament length due to
the stochastic nature of actin polymerization renders the
theoretical analysis of the I-N transition less definitive than
for the simpler monodisperse systems. Recent experimental studies
show that F-actin forms a nematic phase at slightly above 2 mg/ml
protein concentrations
\cite{Kerst90,Coppin92,Furukawa93,Suzuki91}, and the onset actin
concentration for the I-N transition is inversely proportional to
the average filament length, $\lbar$ \cite{Coppin92,Suzuki91}.
There have also been bulk measurements suggesting a concentration
range in which partially aligned domains exist \cite{Suzuki91}.
The goal of our study was set to determine the I-N transition
phase diagram by measuring local alignment of actin filaments,
the co-existing domains of aligned filaments (N) and filaments
with random orientations (I), all in close correlation with
variation of local protein concentrations, \textit{c}. Through
these measurements, we discovered surprising features suggesting
that the I-N transition of F-actin is continuous as a function of
\textit{c}.

For a sample of an aligned array of rodlike filaments, the
difference in index of refraction between the direction of the
alignment and the direction perpendicular to it can be measured
optically. This difference, known as the optical birefringence,
$\Delta n$, is directly proportional to the order parameter of a
uniaxial nematic suspension, $S = \int d\Omega\; f(\theta)
P_{2}(\theta)$, where $f(\theta)$ is the orientational
distribution function and $P_{2}(\theta) =
\left(3\cos^{2}{(\theta)} - 1\right)/2$ is the 2nd Legendre
polynomial. In this report, the birefringence measurements were
performed on a Nikon TE-300 microscope, equipped with a Cambridge
Research Inc (CRI) Polscope package (Cambridge, MA). The Polscope
software is capable of determining $\Delta n$ at each pixel
position, thus reporting local $\Delta n$, as well as the
directions of slow axis throughout the sample
\cite{Oldenbourg95}. In contrast, other birefringence measurement
techniques average over a bulk sample, and are prone to large
variability due to cancellation of random orientations of
separately aligned domains. The measured values by the Polscope
can also be averaged over all pixel values for the entire field
of view, free of the cancellation effect due to different
alignment directions at different positions.

\begin{figure}
\includegraphics[width=3.0truein]{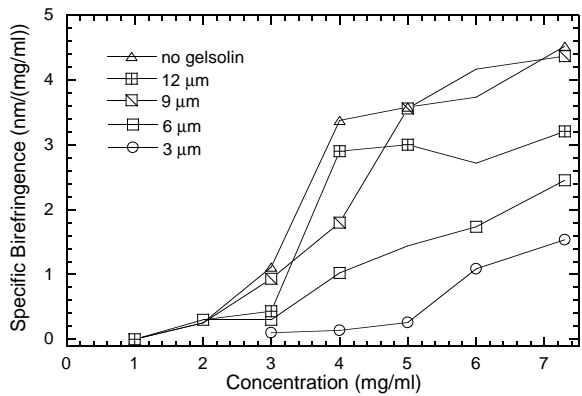}
\caption{Average specific birefringence, $\Delta n/c$, plotted
for F-actin in the concentration range of 1.0-7.3 mg/ml, and of 5
different  $\lbar$. The data show the region of I-N transition
for F-actin.} \label{fig01}
\end{figure}

Fig.~\ref{fig01}
shows the birefringence of F-actin as a function of \textit{c}
and $\lbar$, measured using the Polscope technique. Rectangular
capillary tubes of 0.4 mm thickness and 8 mm width were used to
observe the alignment of F-actin in large domains. After
injecting an actin solution of the desired concentration into the
capillary tube immediately following the addition of 50 mM KCl
and 2 mM $\textrm{MgCl}_{2}$ to start polymerization, both ends
were sealed with melted plastic. The sample was aligned by upside
down inversions using a table top centrifuge. Repeated spins
showed that the average birefringence of the sample reached a
saturation value after two or three inversions (data not shown).
Our measurements confirm the previous findings that the threshold
concentration of the I-N transition is inversely related to
$\lbar$ [11, 13]. The specific birefringence, $\Delta n/c$,
increases with the \textit{c} and reaches a saturated level at the
long filament conditions. Further increase in \textit{c} increased
$\Delta n$ but not $\Delta n/c$, suggesting that the entire
sample was in the nematic state. Actin samples of shorter $\lbar$
did not reach the saturation level for $\Delta n/c$, suggesting
that either the filaments were only weakly aligned, or that the
measured value was the weighed average of co-existing domains.
Variation of birefringence with position was observed, but usually
over relatively large regions with no sharp boundaries. No clearly
separated domains of either I or N phase were detected based on
the birefringence measurements.

\begin{figure*}
\includegraphics[height=4truein]{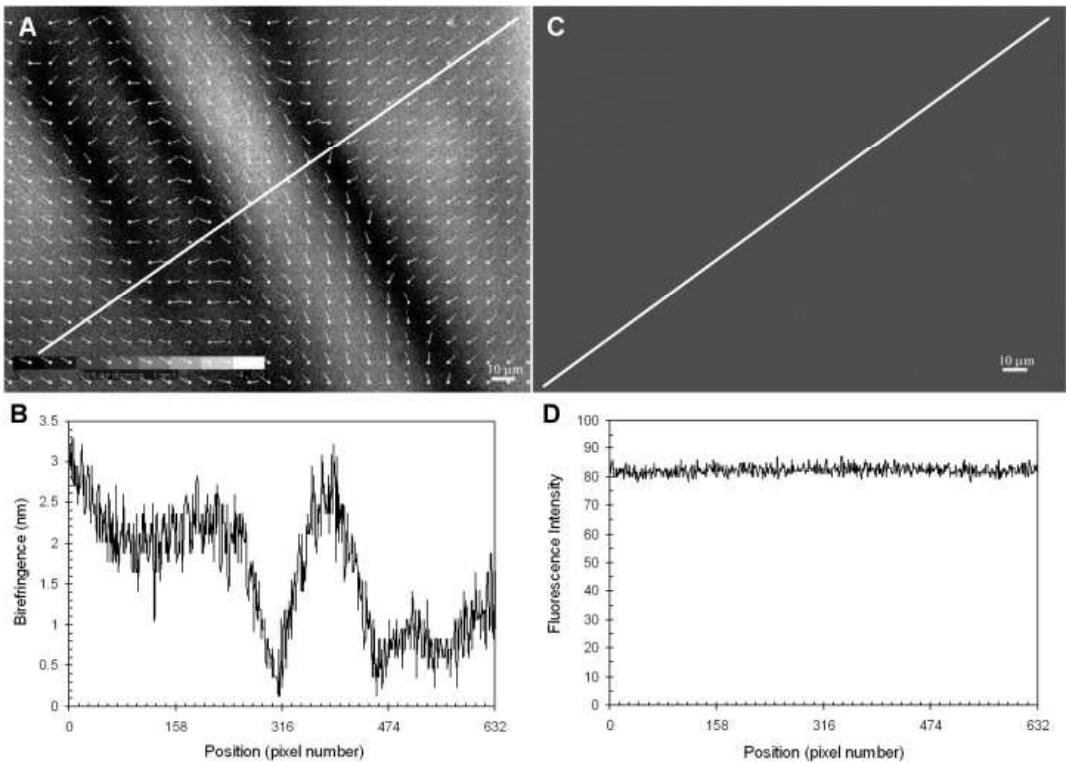}
\caption{\textbf{A}. Birefringence values of 7.3 mg/ml F-actin
with $\lbar = 3\mu$m. The gray scale of 0--4 nm covers the range
of $\Delta n$. \textbf{B}. A line plot of $\Delta n$ along the
diagonal cut, shown from the top right to the bottom left.
\textbf{C}. Fluorescence image of the same region. The measured
intensity values varied from 82 to 85 only, over the camera range
of 0--255. \textbf{D}. A line plot diagonally across the image,
showing little variation in intensity value. The sample thickness
is $10~\mu$m.} \label{fig02}
\end{figure*}

Co-existing I-N domains were expected theoretically to contain
filaments of different concentrations, which is the basic
criterion for a 1st order phase transition of a rodlike
suspension. In order to detect the concentration difference
between co-existing I-N domains of F-actin, TRITC- phalloidin
(Sigma, St. Louis, MO) was added into F-actin to a ratio of 1/1000
and the intensity of fluorescence provided a quantitative report
of concentration variation, especially in regions of the capillary
where local birefringence varied. Measurements of birefringence
and fluorescence were performed for F-actin samples in order to
correlate variations of both parameters. Fig.~\ref{fig02}
displays a set of combined measurements in one representative
condition of a 7.3~mg/ml F-actin sample of $\lbar = 3\mu$m. It is
shown in Fig.~\ref{fig01} that F-actin under this condition is in
the I-N transition region. Therefore, one would expect to have
I-N co-existence based on the theoretical predictions of a 1st
order transition. The $\Delta n$  in the particular region shows
alternating stripes of high and low values, a phenomenon reported
earlier \cite{Coppin92,Suzuki91}. Such a zebra pattern with white
stripes and black grooves appeared occasionally in both thick (up
to 0.4 mm thickness using a rectangular capillary) and thin
samples ($10~\mu$m, for example, as shown in Fig.~\ref{fig02}).
Due to the continuous variation in both $\Delta n$ and direction
of the slow axis over the entire region, the dark and bright
areas do not correspond to separate I-N domains. The black grooves
are not disclination lines \cite{Chandrasekhar86}. In most cases
of our observations for the nematic samples of $\lbar \geq 3\mu$m,
zebra patterns were either not formed, or too fuzzy to be
discerned. Instead, the birefringence pictures appeared rather
uniform, with few line defects. Disclinations were observed less
frequent than once per hundreds of sample slides.

The corresponding fluorescence image (Fig.~\ref{fig02}, C and D)
shows constant \textit{c}  over the entire region of this sample.
The same result of uniform concentration holds for weakly
birefringent regions of less defined patterns, as well (data not
shown). In all actin concentrations up to 7.3 mg/ml and $\lbar
\geq 3\mu$m, we saw no discernable domain boundaries of
co-existence based on the birefringence data. Instead, a blending
of domains of comparable levels of alignment was occasionally
observed, which led to either zebra patterns or other variants.
In all cases, \textit{c} was measured to be strictly uniform,
forcing us to conclude that no co-existence occurred to F-actin
spanning the region of I-N phase transition.

\begin{figure}
\includegraphics[width=3.0truein]{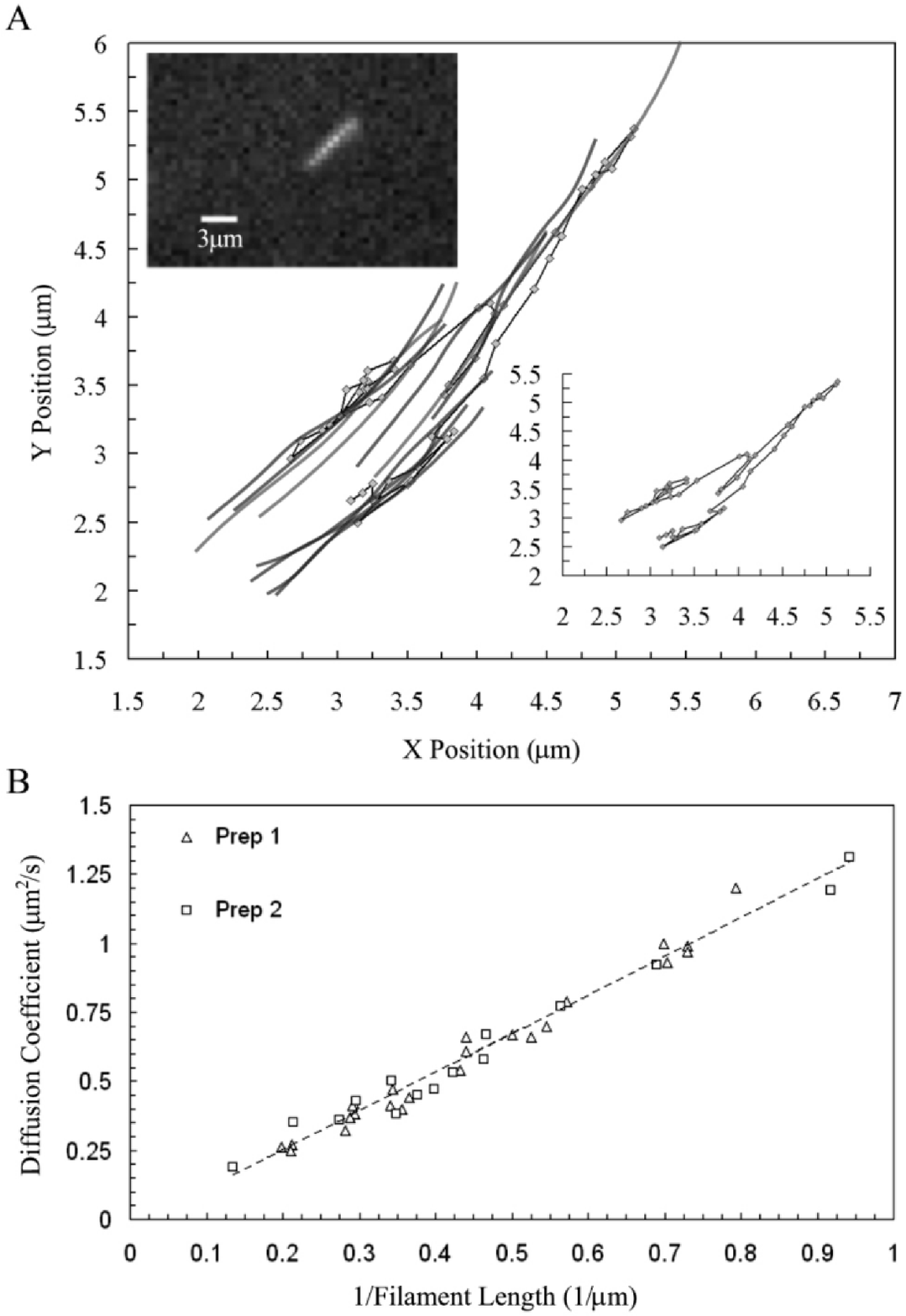}
\caption{Measurement of filament diffusion by image recording.
\textbf{A}. Configuration and traces of one labeled filament in
the nematic matrix. Left inset: a typical frame. Right inset:
trajectory of the filament without overlay of the filament images.
\textbf{B}. Diffusion coefficient determined for various filament
length and two F-actin preparations. The fit line shows $D \propto
1/L$.}\label{fig03}
\end{figure}

Additional experiments were performed to examine two possible
causes for the lack of concentration variation in the I-N
transition of F-actin. First, we suspected that the F-actin
network at several mg/ml might be weakly crosslinked
\cite{Tang99}. As a result, the polymerizing actin solution may
first form a nematic phase when $\lbar$  reaches the value
required by the excluded volume effect. Further polymerization
and/or annealing by the crosslinking effect may lead to gelation
so that the network is unable to achieve its true thermodynamic
equilibrium, with well separated I-N domains in co-existence. A
second possibility is that the extremely long F-actin tend to get
sufficiently entangled and thus kinetically trapped in a gel-like
or glassy state with local orientational order but no global
phase equilibrium. We address these concerns by mixing a small
number of labeled filaments with 7 mg/ml unlabeled F-actin, both
of $\lbar = 3\mu$m, and show that the labeled filaments undergo
rapid thermal motions. Fig.~\ref{fig03}
shows a time sequence of one filament in a nematic domain by
overlaying images with 0.5 sec intervals. The filaments diffuse
preferentially along their axis. The diffusion coefficient of
each labeled filament was determined by calculating the
$\langle\Delta x^{2}\rangle$ over a movie sequence of 50--150
frames. The translational diffusion of labeled filament along its
axis is found to be $1.30/\mathrm{L}~(\mu m^{2}/\textrm{sec})$,
where L, omitting its unit of $\mu$m, is the length of the
diffusing filament. The pre-factor of 1.30 obtained from the fit
in Fig.~\ref{fig03} is smaller than the theoretical value of 3.90
for the translational diffusion along filament axis in the dilute
limit. This reduction by a factor of 3 is not surprising since
F-actin are highly entangled at a few mg/ml. In contrast, the
lateral and rotational diffusion coefficients are suppressed by
factors of at least 10 and 100, respectively, although we were
unable to obtain these small values limited by the image
resolution.

\begin{figure}
\includegraphics[width=3.0truein]{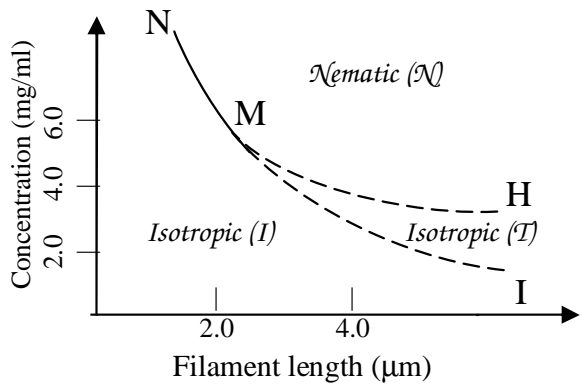}
\caption{Sketch of a proposed phase diagram for the I-N transition
of long and stiff rods. Each phase region follows what is shown in
Fig.~2 of ref.~\cite{Lammert93}. The nematic interaction (J) and
the defect suppression (K) parameters of ref.~\cite{Lammert93} are
mapped to the concentration and filament length, respectively.}
\label{fig04}
\end{figure}

We propose that the I-N transition of F-actin of $\lbar \geq
3\mu$m corresponds to the continuous transition between a
topologically ordered isotropic phase (T) and a weakly aligned
nematic phase (N), as predicted theoretically by Lammert, Rokhsar
\& Toner (LRT) \cite{Lammert93}. First, the suspension of long
(aspect ratio over 300) and stiff (persistence length $15~\mu$m
\cite{Gittes93}) actin filaments satisfy one key criterion of
defect suppression, which according to the LRT theory renders the
transition continuous. The disclination lines are energetically
costly for F-actin, as the bending elastic constant
($\mathrm{K_{3}}$) is far greater than splay ($\mathrm{K_{1}}$)
and twist ($\mathrm{K_{2}}$) for long and stiff rods
\cite{Lee86}. Indeed, few line defects and nearly no
disclinations were observed in F-actin samples. Second,
suppression of rotational diffusion likely leads to breaking the
inversion symmetry of long filaments in the concentrated F-actin.
Therefore, vectorial directionality can be assigned to individual
filaments, and in this context, the isotropic state becomes
topologically ordered. A schematic phase diagram for F-actin as
shown in Fig.~\ref{fig04},
which translates the main predictions of the LRT theory, as
depicted in Fig.~2 of ref.~\cite{Lammert93}.

We now briefly discuss two other features that affect the I-N
transition: the filament flexibility and the polydispersity of
length. The flexibility is known to suppress the concentration gap
between co-existing I-N domains. For instance, such a gap for the
I-N transition of the semiflexible fd/M13 viruses is reduced to
about 10\% of the up limit concentration of the isotropic phase
\cite{Tang95}, down from over 30\% predicted by the Onsager
theory for rigid rods \cite{Onsager49}. The fd or M13 virus has
about the same diameter as F-actin, but a persistence length of
$2.2~\mu$m \cite{Tang95}, much shorter than that of F-actin
(15--$20~\mu$m) \cite{Gittes93}. Therefore, the effect of limited
flexibility for F-actin is insufficient to lead to disappearance
of the concentration gap characteristic of a 1st order transition
for most polymer liquid crystals \cite{Ciferri82}. While
polydispersity has been speculated as a possible cause of
smearing the I-N phase boundary, calculations for bidispersed
rods suggest not narrower but wider co-existing regions
\cite{Vroege93}. Experimental study of polydisperse systems of
short rods suggests a wider co-existence region for the I-N
transition \cite{Buitenhuis95}.

Dogic and Fraden recently attempted to detect co-existence for the
I-N transition of pf1, another filamentous bacteriophage of
$2~\mu$m length. Similar to our findings for the long F-actin,
they observed no domain separation for the I-N transition (Dogic
and Fraden, private communications). Since the pf1 virus has
similar diameter and persistent length to fd/M13, but is twice as
long, the different behavior between their I-N transitions
implies that even the $2~\mu$m filaments are perhaps too long to
lead to domain separation. The similar behavior of pf1 to F-actin
also supports the argument that polydispersity is unlikely to
account for the observed continuous transition for F-actin, since
pf1 viruses are monodisperse.

In summary, the I-N transition for F-actin of $\lbar \geq 3\mu$m
is continuous in alignment and protein concentration. This
experimental finding, and more importantly, its proposed relevance
to the LRT theory, will likely provoke further query, such as a
search for the I-T transition closely related to the T-N
transition and the resultant critical fluctuations.

\begin{acknowledgments}
We appreciate helpful discussions with Bob Pelcovits, John Toner,
James Swihart, Seth Fraden, and Zvonimir Dogic. This work is
supported by NSF DMR 9988389. Jorge Viamontes is a GAANN fellow,
supported by the US Department of Education.
\end{acknowledgments}


\begin{thebibliography}{99}
\bibitem{Sato97} Sato, T., Y. Jinbo, and A. Teramoto. Macromolecules, 1997. 30: 590.
\bibitem{Buxbaum87} Buxbaum, R.E., et al. Science, 1987. 235: 1511.
\bibitem{Kerst90} Kerst, A., et al. Proc.\ Natl.\ Acad.\ Sci., 1990. 87: 4241.
\bibitem{Tang95} Tang, J.X. and S. Fraden. Liquid Crystals, 1995. 19: 459.
\bibitem{Onsager49} Onsager, L. Ann. NY Acad. Sci., 1949. 51: 627.
\bibitem{Flory69} Flory, P.J., Statistical Mechanics of Chain
Molecules. 1969, Interscience Publishers, New York.
\bibitem{deGennes94} de Gennes, P.G. and J. Prost, The Physics of
Liquid Crystals. 1994, Clarendon: Oxford.
\bibitem{Lammert93} Lammert, P.E., D.S. Rokhsar, and J. Toner. Phys.\ Rev.\ Lett., 1993. 70:
1650. Phys.\ Rev.\ E, 1995, 52: 1778 and 1801.
\bibitem{Kas95} Kas, J., et al. Biophys.\ J., 1995. 70: 609.
\bibitem{Coppin92} Coppin, C. and P. Leavis. Biophys.\ J., 1992. 63: 794.
\bibitem{Furukawa93} Furukawa, R., R. Kundra, and M. Fechheimer. Biochemistry, 1993. 32: 12346.
\bibitem{Suzuki91} Suzuki, A., T. Maeda, and T. Ito. Biophys. J., 1991. 59: 25.
\bibitem{Oldenbourg95} Oldenbourg, R. and G. Mei. J. of Microscopy, 1995. 180: 140.
\bibitem{Chandrasekhar86} Chandrasekhar, S. and G.S. Ranganath. Adv.\ in Phys., 1986. 35: 507.
\bibitem{Tang99} Tang, J.X., et al. Biophys.\ J., 1999. 76: 2208.
\bibitem{Lee86} Lee, S.D. and R.B. Meyer. J.\ Chem.\ Phys., 1986. 84: 3443.
\bibitem{Gittes93} Gittes, F., et al. J.\ Cell Bio., 1993. 120: 923.
\bibitem{Ciferri82} Ciferri, A., W.R. Krigbaum, and R.B. Meyer,
eds. Polymer Liquid Crystals. 1982, Academic Press: New York.
\bibitem{Vroege93} Vroege, G.L. and H.N.M. Lekkerkerker. J.\ Phys.\ Chem., 1993. 97: 3601.
\bibitem{Buitenhuis95} Buitenhuis, J., et al. J. of Colloid and Interface Science, 1995. 175: 46.
\end{thebibliography}
\end{document}